# Nonlinear refraction in $CH_3NH_3PbBr_3$ single crystals

Christian Kriso,[1,*] Markus Stein,[1] Tobias Haeger,[2] Neda Pourdavoud,[2] Marina Gerhard,[1] Arash Rahimi-Iman,[1] Thomas Riedl,[2] and Martin Koch[1]

[1]Department of Physics and Material Sciences Center, Philipps-Universität Marburg, Renthof 5, 35032 Marburg, Germany

[2]Institute of Electronic Devices, University of Wuppertal, Rainer-Gruenter-Strasse 21, 42119 Wuppertal, Germany

[*]christian.kriso@physik.uni-marburg.de

**Hybrid lead halide perovskites, such as $CH_3NH_3PbX_3$ (X=I, Br), are direct gap semiconductors that offer many superior optoelectronic properties combined with extremely simple solution-processing fabrication methods. This makes them very attractive for use in applications like solar cells or light-emitting devices. Recently, also their nonlinear optical properties have received increased attention due to reports of high nonlinear refraction in thin films and nanoparticles. However, understanding of the underlying mechanisms is poor and limited by the lack of knowledge of fundamental parameters like the nonlinear refractive index of the bulk material. Here, we measure both nonlinear absorption and nonlinear refraction in a $CH_3NH_3PbBr_3$ single crystal using the Z-scan technique with femtosecond laser pulses. At 1000 nm, we obtain values of 5.2 cm/GW and $9.5 \cdot 10^{-14}$ cm$^2$/W for nonlinear absorption and nonlinear refraction, respectively. Sign and magnitude of the observed refractive nonlinearity are reproduced well by the two-band model. To our knowledge, these measurements mark the first characterization of nonlinear refraction in any metal halide perovskite single crystal and thus will serve as an important reference for assessing the potential of this emerging material class for nonlinear optical applications.**

Metal halide perovskites have seen a tremendous interest recently due to the demonstration of high efficiency solar cells with thin film processed devices offering the prospect to revolutionize the field of photovoltaics in both fabrication costs and efficiency.[1] Among strong efforts to further improve the understanding of the intricate photophysics and to exploit their highly luminescent properties for light-emitting devices and lasers,[2,3] metal halide perovskites have gained considerable interest in exploring their nonlinear optical properties.[4–7] Perovskite thin films,[4,5,8] nanocrystals,[8,9] and two-dimensional perovskites[10] have shown promising nonlinear optical properties in terms of absolute values as well as in their figure of merit (FOM). The FOM relates the magnitude of nonlinear refraction to the amount of nonlinear absorption and thus indicates how well a material is suited for applications such as nonlinear optical switching. In this context, the question arises whether the so far reported promising nonlinear optical properties of perovskites are the result of the particular nanostructure of the samples investigated, overestimation due to scattering and thermal effects or whether they represent an intrinsic property of the material class. Up to now, no reports exist on the intrinsic refractive nonlinearity of any metal halide perovskite single crystal to shed light on this fundamental question.[6,7] (Note that the value of nonlinear refraction for MAPbI3 single crystals listed in Ferrando *et al.* could not be found in the given or any other reference such that we think that Ferrando *et al.* must have confused it with another parameter.[6]) Metal halide perovskites single crystals have been used to probe the ultimate limit of carrier transport and have shown to exhibit extraordinary low trap-state densities and large diffusion lengths.[11,12] Knowledge about the refractive index nonlinearity in perovskite single crystals would help to separate the intrinsic nonlinearity of the material system from possible enhancement effects of the nonlinearity in perovskite nanostructures due to confinement or carriers.[13] This is also of particular relevance as grain boundaries and other sample imperfections can lead to strong scattering in Z-scan experiments, which cast serious doubt on the validity of some measurements and their extracted values of the nonlinear refractive index.[4,6,8]

Here, we measure the nonlinear refractive index of a $CH_3NH_3PbBr_3$ (or $MAPbBr_3$) single crystal with the Z-scan method and compare the obtained value to reports of nonlinear refraction in various metal halide perovskite nanostructures. Apparently, some of the values reported for nanostructured samples substantially deviate from the single crystal data. As such, our results raise the question to what extent nonlinear optical properties reported for nanoparticles can be indeed attributed to confinement or carrier effects in the nanostructures and to what extent they may be the result of an overestimation due to the limitations of the Z-scan technique when using highly polycrystalline samples.

**Results and Discussion**

To verify the properties of the synthesized $MAPbBr_3$ single crystals, we initially determined the linear absorbance and the one-photon induced PL (in reflection geometry) of the single crystal $MAPbBr_3$ sample excited by a frequency-doubled, mode-locked Ti:Sapphire laser at 440 nm. The optical characteristics agree well with the results of Wenger *et al.* and are displayed in **Figure 1a**.[14] **Figure 1b** shows representative XRD data of the synthesized $MAPbBr_3$ single crystals, which are in good agreement with reference data.[15] In addition, current-voltage measurements of $MAPbBr_3$ single crystals synthesized in the same manner show very low trap state densities (see **Figure S1** in the supporting information) further demonstrating the high quality of the crystals.

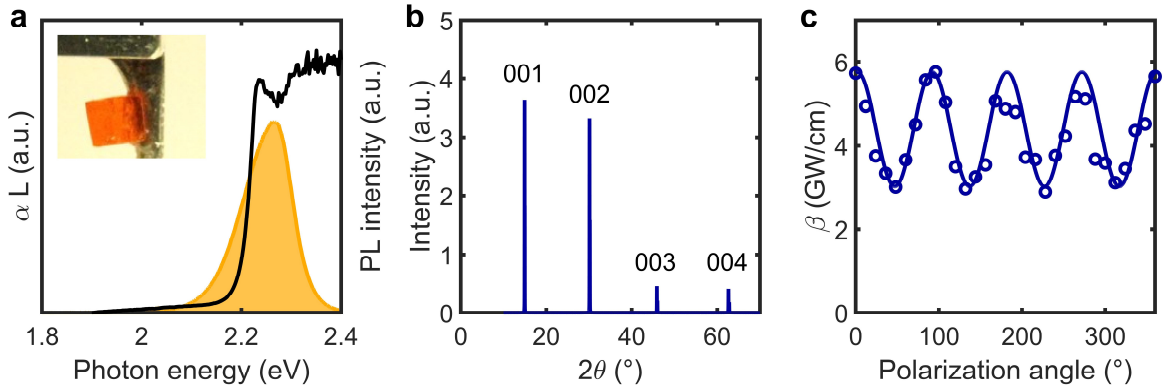

**Figure 1**: (a) Absorbance (black line) and PL (yellow area) of the $MAPbBr_3$ single crystal sample when excited with a pulsed laser at 440 nm. The inset shows a photo of the sample attached to the sample holder. (b) Representative XRD data of the fabricated single crystals. (c) Two-photon absorption coefficient of the sample as a function of incident polarization of the probe laser at 880 nm. The solid line represents the theoretical model.

In addition to that, we measured the polarization-resolved two-photon absorption coefficient with the Z-scan technique with the same probe laser at 880 nm, which is shown in **Figure 1c**. The obtained polarization dependence complies well with the expected behavior for a cubic crystal structure,[16]

$$\beta(\theta) = A\{1 + 2\sigma[\sin^4(\theta + \phi) - \sin^2(\theta + \phi)]\}, \quad (1)$$

where the parameters $A$ = 5.75 GW/cm, $\sigma$ = 0.95 and $\phi$ = 2 yield good correspondence with the measured data. The large value of the anisotropy parameter $\sigma$ thereby highlights the highly crystalline structure of the sample compared to other values reported in literature.[17,18] For a comparison of the absolute magnitude of nonlinear absorption, we refer to a later section in the text, where we discuss nonlinear absorption measurements at 1000 nm with a kHz laser system.

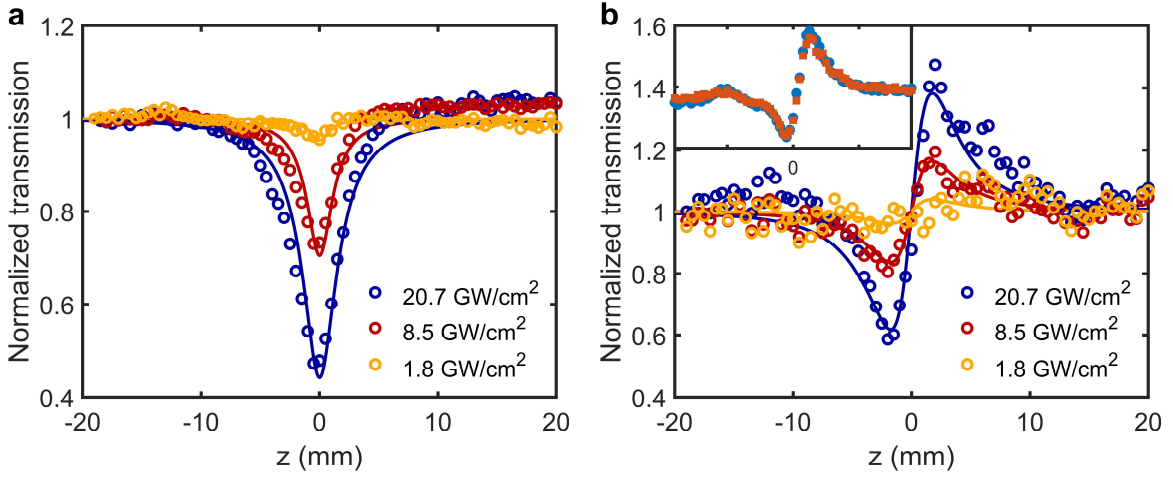

**Figure 2:** Z-scan measurements of (a) nonlinear absorption and (b) nonlinear refraction at 1000 nm for different probe intensities at 1 kHz repetition rate. The solid lines represent fit functions that are used to extract nonlinear absorption and refraction. The inset in (b) displays Z-scans measured with the same peak intensity but different repetition rates of the probe laser (blue circles: 1 kHz, orange squares: 100 Hz).

**Figure 2** shows exemplary Z-scan measurements of nonlinear absorption and nonlinear refraction for different incident intensities at a probe wavelength of 1000 nm. For fitting the open-aperture scans of **Figure 2a,** we use the model of Sheik-Bahae et al.,[19]

$$T(z) = \frac{1}{\sqrt{\pi}q_0(z)} \int_{-\infty}^{+\infty} \ln[1 + q_0(z)e^{-\tau^2}] d\tau, \qquad (2)$$

where $q_0(z)=\beta I_0 L_{eff}/(1+z^2/z_0^2)$ with $\beta$ being the two-photon absorption coefficient, $I_0$ the on-axis peak intensity of the probe laser pulse, $L_{eff}=(1-exp(-\alpha L))/\alpha$ the effective sample length with $L$ being the physical sample length, $\alpha$ the linear absorption coefficient and $z_0$ the Rayleigh length of the focused beam. The total absorption coefficient is then given by $\alpha+\beta I_0$. At 1000 nm, nonlinear refraction is positive as indicated by the valley preceding the peak of the Z-scan around the focus in **Figure 2b**. These Z-scan measurements of nonlinear refraction are obtained by dividing the partially closed aperture scan by the open aperture scan and normalizing it to transmission one at $z$ positions far from focus. For fitting, we use a model for optically thick samples,[20]

$$T(z) = 1 + \Delta\Phi_0 F(x,l), \qquad (3a)$$

$$F(x,l) = \frac{1}{4}\ln\left(\frac{\left[\left(x+\frac{l}{2}\right)^2+1\right]\left[\left(x-\frac{l}{2}\right)^2+9\right]}{\left[\left(x-\frac{l}{2}\right)^2+1\right]\left[\left(x+\frac{l}{2}\right)^2+9\right]}\right), \qquad (3b)$$

with $x=z/z_0$, $l=L/z_0$ and $z_0 =n_0\pi w_0^2/\lambda$ being the Rayleigh range within the sample. Here, $n_0$ is the linear refractive index, which has been taken as 2 according to Park et al.,[15] and $w_0$ is the beam waist at focus which is derived as 18±2 μm. $\Delta\Phi_0 =kn_2 I_0 L_{eff}$ is the on-axis nonlinear optical phase shift where $k$ is the wave number and $n_2$ is the nonlinear refractive index. The total refractive index is $n_0+n_2 I_0$. The inset in **Figure 2b** displays two Z-scans at the same peak intensity but at different repetition rates (100 Hz and 1000 Hz) to assess possible thermal effects. Since no significant deviation of the two measurements is observed, we rule out any thermal effects on the measurements and perform them at 1000 Hz. From the above fit functions we can deduce the normalized nonlinear absorption $q_0 = q_0(0)$ and nonlinear on-axis phase shift $\Delta\Phi_0$, which we plot in **Figure 3a** and **b** as a function of on-axis peak intensity, respectively. Both quantities follow approximately a linear trend with increasing peak intensity confirming the presence of a third-order nonlinearity. The stronger fluctuations of the data points around the linear fit function in **Figure 3b** compared to **Figure 3a** are attributed to the fact that the Z-scan measurements in **Figure 2b** contain more noise than the open aperture scans due to the smaller absolute power levels present in these measurements and the additional division by the open aperture scan.

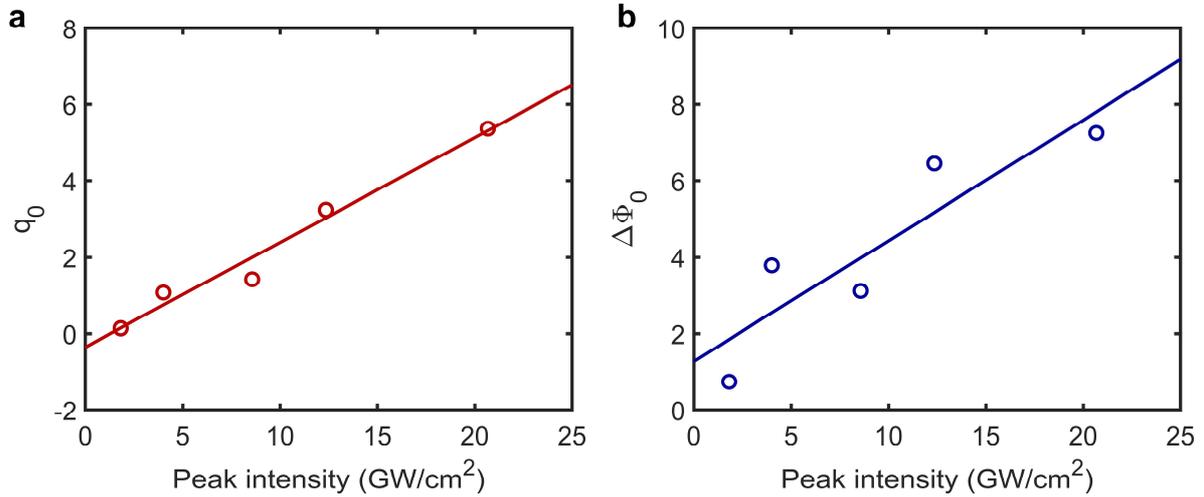

**Figure 3**: (a) Normalized nonlinear absorption and (b) on-axis nonlinear phase shift showing a linear trend with increasing peak intensity, thus confirming a third-order nonlinearity.

From the slope of the interpolation functions in **Figure 3a** and **b** we deduce the nonlinear absorption and refraction coefficients, $\beta$ and $n_2$, respectively. Main error source in their determination is the uncertainty of the peak power density at focus, which we estimate by considering the inaccuracies in the pulse length and beam width determination and the average power fluctuations. By fitting the data points as a function of the upper and lower limit of the intensity scale, that considers these variations, we retrieve the uncertainty of the nonlinear optical coefficients. We treat scattering at the surfaces and within the sample like linear absorption by defining a linear absorption coefficient $\alpha$ as done by Saouma et al.[18] As we measure a total transmission of $T^2 \exp(-\alpha L)$ = 0.52, but, considering only the reflection at the interfaces, total transmission should amount to 0.79, we incorporate all additional losses into $\alpha$. Then, the effective length $L_{eff}$ yields 530 µm while the physical length $L$ of the sample is 650 µm. As a result, we obtain $\beta$ = 5.2(+2.7/-1.3) cm/GW for nonlinear absorption and $n_2$ = 9.5(+5.0/-2.4)·10$^{-14}$ cm$^2$/W for nonlinear refraction. The asymmetric errors are a result of the above-mentioned extraction procedure, which uses the slope of normalized nonlinear absorption and nonlinear phase shift as function of intensity to extract the nonlinear optical coefficients. We point out that the nonlinear refractive index of any metal halide perovskite single crystal has not been measured before. The value of nonlinear absorption is somewhat smaller than the values obtained by Walters et al. and Saouma et al. for MAPbBr$_3$ single crystal, with 8.6 cm/GW at 800 nm and 9 cm/GW at 1064 nm, respectively.[17,18] However, our value of $\beta$ is well within the range of the polarization-resolved $\beta$ from our preliminary investigation conducted at 880 nm and depicted in **Figure 1c**.

Non-resonant, nonlinear absorption and refraction is often modeled by the two-band model, which relates the nonlinear optical absorption via Kramers-Kronig relations to nonlinear refraction.[21] Taking the model for two-photon absorption,

$$\beta(\omega) = K \frac{\sqrt{E_p}}{n_0^2 E_g^3} F_2\left(\frac{\hbar\omega}{E_g}\right), \qquad (4)$$

and for nonlinear refraction,

$$n_2(\omega) = K \frac{\hbar c \sqrt{E_p}}{n_0^2 E_g^4} G_2\left(\frac{\hbar\omega}{E_g}\right), \qquad (5)$$

with the material independent parameters $K$=3100, the Kane parameter $E_p$=21 eV and the dispersion functions $F_2(x)$ and $G_2(x)$ (with $x$ being the photon energy normalized to the band gap), as specified in Sheik-Bahae et al.,[21] and only using the material specific parameters for the band gap, $E_g$ = 2.2 eV, and the refractive index, $n_0$ = 2, we obtain $\beta$ = 8.2 cm/GW and $n_2$ = 8.1·10$^{-14}$ cm$^2$/W at 1000 nm. These values are remarkably close to our experimentally obtained results including the same positive sign for nonlinear refraction.

| Structure | Material | Band gap (eV) | Wavelength (nm) | Pulse properties | $\beta$ (cm/GW) | $n_2 \cdot 10^{-14}$ (cm$^2$/W) | Ref. |
|---|---|---|---|---|---|---|---|
| Single crystal | MAPbBr$_3$ | 2.2 | 1000 | 70 fs, 1 kHz | 5.2 | 9.5 | This work |
| Thin film | MAPbBr$_3$ | 2.21 | 1064 | 1 ns, 20 kHz | - | 11000-35000 | 8 |
| Thin film | MAPbI$_3$ | 1.61 | 1028 | 200 fs, 1kHz | -500 (SA)-272000 | 210-1900 | 4 |
| Thin film | MAPbI$_3$ | 1.61 | 1064 | 40 ps, 10 Hz | -2250 (SA) | 3740 | 5 |
| Nanocrystals | MAPbBr$_3$ | 2.38 | 800 | 130 fs, 76 MHz | 4.1 | (-40.1) | 9 |
| Nanocrystals | CsPbBr$_3$ | 2.36-2.38 | 600-800 | 50 fs, 1 kHz | 0.5-180 | 1-290 | 22 |

**Table 1:** Comparison of the MAPbBr$_3$ single crystal nonlinear optical properties to those of several other perovskite nanostructures from literature. SA stands for saturable absorption.

For comparison, GaAs shows roughly three times larger values for the nonlinear coefficients, with $\beta$ = 26 cm/GW and $n_2$ = -31·10$^{-14}$ cm$^2$/W at 1064 nm.[23] The FOM = $|1/\lambda \cdot n_2/\beta|$ is approximately the same, while GaAs at this wavelength exhibits 0.11 we obtain here 0.1 for MAPbBr$_3$.

More importantly, we can compare our results to previous measurements of nonlinear refraction in perovskite thin films and nanocrystals (**Table 1**). So far, for MAPbBr$_3$ there have only been reports about measurements on thin films and colloidal nanocrystals.[8,9] For thin films, Suarez et al. obtained values of $n_2$ between 1.1·10$^{-9}$ and 3.5·10$^{-9}$ cm$^2$/W at an excitation wavelength of 1064 nm with ns long pulses.[8] This is five orders of magnitude higher than the values obtained here for the bulk material and demands for more detailed investigations of the mechanisms that enhance the nonlinear optical response in thin films so drastically. In addition, the possible influence of the pulse length in the measurement needs further clarification.

Lu et al. claim negative nonlinear refraction of MAPbBr$_3$ nanoparticles of -4.01·10$^{-13}$ cm$^2$/W at 800 nm probed by femtosecond pulses.[9] However, we believe that the use of a 76 MHz repetition rate laser might have obscured their measurements of an ultrafast Kerr nonlinearity as we also measured a negative nonlinearity when using an 80 MHz repetition rate laser in preliminary investigations. However, by resolving the temporal dynamics at the μs scale we were able to trace this back to a rather slow thermal nonlinearity as shown in the supporting information in **Figure S3**.

For MAPbI$_3$ thin films, both Kalanoor *et al.* and Zhang *et al.* report large values of nonlinear refraction, 1.9·10$^{-11}$ cm$^2$/W with fs pulses and 3.7·10$^{-11}$ cm$^2$/W with ps pulses, respectively.[4,5] There, even for non-resonant probing close to 1000 nm large saturable absorption is observed in the open-aperture Z-scan measurements, which is attributed to the presence of sub-gap states. Thus, the observed enhancement of the refractive nonlinearity in these thin films can most likely be explained with carrier effects. Time-resolved measurements of the nonlinear refractive index could provide further insight into this phenomenon.

It is important to note that substantially smaller numbers for nonlinear refraction of CsPbBr$_3$ nanocrystals in the range of 1·10$^{-14}$ cm$^2$/W and 290·10$^{-14}$ cm$^2$/W were obtained at wavelengths from 600 nm to 800 nm if lower repetition rates of 1 kHz and femtosecond pulses were used.[22]

**Conclusions**

In conclusion, we have measured nonlinear refraction in a MAPbBr$_3$ single crystal. We find a nonlinear refractive index of 9.5·10$^{-14}$ cm$^2$/W at 1000 nm, which is in good agreement with the two-band model. These results show that the significantly larger values of nonlinear refraction reported in the literature for thin films and nanoparticles cannot be explained with an intrinsically high nonlinear refractive index in metal halide perovskites but rather by carrier or confinement effects. Our results provide the first report of nonlinear refraction in a high-quality metal halide perovskite single crystal. We expect that our findings will trigger further investigations of the underlying mechanisms of nonlinear refraction in perovskite single crystals and nanostructures.

**Methods**

**Sample fabrication:**

MAPbBr$_3$ single crystals were grown by the inverse temperature crystallization procedure.[24] Briefly, a 1 M solution of PbBr$_2$ (Sigma Aldrich) and methylammonium bromide (Dyesol) was prepared in anhydrous dimethylformamide (DMF from Sigma-Aldrich) at room temperature. The MAPbBr$_3$ perovskite crystals were grown at 80°C. More details about the growth and characteristics of the MAPbBr$_3$ single crystals can be found in our previous paper.[25] For the Z-scan measurements, the samples were cleaved on both sides to improve the optical quality.

**Current-Voltage traces**

The MAPbBr$_3$ single crystals were cleaved from both sides to achieve a fresh surface. Au electrodes (thickness 150nm) were thermally evaporated onto both fresh surfaces in a vacuum chamber. The I/V characteristics were recorded in darkness using a Keithley 236 source measure unit. The trap density has been derived following the procedure detailed in Saidaminov *et al*.[24]

**Experimental setup:**

For Z-scan measurements of nonlinear refraction, we use a 1 kHz regenerative amplifier system (Spectra-Physics Spitfire Ace), that drives an optical parametric amplifier, which emits pulses centered at 1000 nm. The 1 kHz repetition rate of the probe laser prevents cumulative heating of the sample that otherwise might lead to strong thermal lensing and thus overshadows an ultrafast Kerr nonlinearity. This has been confirmed by reducing the repetition rate with a chopper down to 100 Hz without observing a significant difference in the Z-scan measurement (see the inset of **Figure 2b**). Autocorrelation measurements of the probe beam yield a pulse duration of 70 fs with an estimated uncertainty of ± 5 fs. The probe beam is focused down to a spot size of 18±2µm with a lens of 10 cm focal length. The sample is mounted on a computer-controlled translation stage which moves the sample through the focused beam. After passing through the sample, the beam is divided by a beam splitter in equal parts and recorded by two large area silicon photodiodes, one of which has a partially closed aperture in front of it. Lock-In detection is used for extracting the signal of both photodiodes. To vary the probe intensity, neutral density filters are inserted in front of the setup. An overview of the setup is displayed in **Figure S2** of the supporting information.

The setup for the measurement of the polarization-resolved two-photon absorption as depicted in **Figure 1c** is very similar, but a 80 MHz mode-locked Ti:Sapphire laser at 880 nm was used. The polarization is rotated with a λ/2-plate inserted in front of the focusing lens.

**Competing interests**

The authors declare no conflict of interest.

**Acknowledgements**

We acknowledge the German Federal Ministry for Education and Research (Grant No. 13N13819) and the DFG (Deutsche Forschungsgemeinschaft) (Grants No. HE2698/7-2, RI1551/9-1, RI1551/12-1 and RA2841/1-1) for financial support.

**References**

(1) Green, M. A.; Ho-baillie, A.; Snaith, H. J. The Emergence of Perovskite Solar Cells. *Nat. Photonics* **2014**, *8*, 506–514. https://doi.org/10.1038/nphoton.2014.134.

(2) Tan, Z.-K.; Moghaddam, R. S.; Lai, M. L.; Docampo, P.; Higler, R.; Deschler, F.; Price, M.; Sadhanala, A.; Pazos, L. M.; Credgington, D.; et al. Bright Light-Emitting Diodes Based on Organometal Halide Perovskite. *Nat. Nanotechnol.* **2014**, *9*, 687–692. https://doi.org/10.1038/nnano.2014.149.

(3) Pourdavoud, N.; Haeger, T.; Mayer, A.; Cegielski, P. J.; Giesecke, A. L.; Heiderhoff, R.; Olthof, S.; Zaefferer, S.; Shutsko, I.; Henkel, A.; et al. Room-Temperature Stimulated Emission and Lasing in Recrystallized Cesium Lead Bromide Perovskite Thin Films. *Adv. Mater.* **2019**, *31*, 1903717. https://doi.org/10.1002/adma.201903717.

(4) Kalanoor, B. S.; Gouda, L.; Gottesman, R.; Tirosh, S.; Haltzi, E.; Zaban, A.; Tischler, Y. R. Third-Order Optical Nonlinearities in Organometallic Methylammonium Lead Iodide Perovskite Thin Films. *ACS Photonics* **2016**, *3*, 361–370. https://doi.org/10.1021/acsphotonics.5b00746.

(5) Zhang, R.; Fan, J.; Zhang, X.; Yu, H.; Zhang, H.; Mai, Y.; Xu, T.; Wang, J.; Snaith, H. J. Nonlinear Optical Response of Organic – Inorganic Halide Perovskites. *ACS Photonics* **2016**, *3*, 371–377. https://doi.org/10.1021/acsphotonics.5b00563.

(6) Ferrando, A.; Pastor, J. P. M.; Suarez, I. Toward Metal Halide Perovskite Nonlinear Photonics ́. *J. Phys. Chem. Lett.* **2018**, *9*, 5612–5623. https://doi.org/10.1021/acs.jpclett.8b01967.

(7) Xu, J.; Li, X.; Xiong, J.; Yuan, C.; Semin, S.; Rasing, T. Halide Perovskites for Nonlinear Optics. *Adv. Mater.* **2019**, *1806736*. https://doi.org/10.1002/adma.201806736.

(8) Suarez, I.; Valles-Pelarda, M.; Guadron-Reyes, A. F.; Mora-Sero, I.; Ferrando, A.; Michinel, H.; Salgueiro, J. R.; Pastor, J. P. M. Outstanding Nonlinear Optical Properties of Methylammonium- and CsPbX3 ( X = Br , I , and Br – I ) Perovskites : Polycrystalline Thin Films and Nanoparticles. *APL Mater.* **2019**, *7*, 041106. https://doi.org/10.1063/1.5090926.

(9) Lu, W.; Chen, C.; Han, D.; Yao, L.; Han, J.; Zhong, H.; Wang, Y. Nonlinear Optical Properties of

Colloidal CH3NH3PbBr3 and CsPbBr3 Quantum Dots : A Comparison Study Using Z-Scan Technique. *Adv. Opt. Mater.* **2016**, *4*, 1732–1737. https://doi.org/10.1002/adom.201600322.

(10) Saouma, F. O.; Stoumpos, C. C.; Wong, J.; Kanatzidis, M. G.; Jang, J. I. Selective Enhancement of Optical Nonlinearity in Two-Dimensional Organic-Inorganic Lead Iodide Perovskites. *Nat. Commun.* **2017**, *8*, 742. https://doi.org/10.1038/s41467-017-00788-x.

(11) Shi, D.; Adinolfi, V.; Comin, R.; Yuan, M.; Alarousu, E.; Buin, A.; Chen, Y.; Hoogland, S.; Rothenberger, A.; Katsiev, K.; et al. Low Trap-State Density and Long Carrier Diffusion in Organolead Trihalide Perovskite Single Crystals. *Science* **2015**, *347*, 519–522. https://doi.org/10.1126/science.aaa2725.

(12) Dong, Q.; Fang, Y.; Shao, Y.; Mulligan, P.; Qiu, J.; Cao, L.; Huang, J. Electron-Hole Diffusion Lengths >175 Mm in Solution-Grown CH3NH3PbI3 Single Crystals. *Science* **2015**, *347*, 967–970. https://doi.org/10.1126/science.aaa5760.

(13) Christodoulides, D. N.; Khoo, I. C.; Salamo, G. J.; Stegeman, G. I.; Van Stryland, E. W. Nonlinear Refraction and Absorption: Mechanisms and Magnitudes. *Adv. Opt. Photonics* **2010**, *2*, 60–200. https://doi.org/10.1364/AOP.2.000060.

(14) Wenger, B.; Nayak, P. K.; Wen, X.; Kesava, S. V; Noel, N. K.; Snaith, H. J. Consolidation of the Optoelectronic Properties of CH3NH3PbBr3 Perovskite Single Crystals. *Nat. Commun.* **2017**, *8*, 590. https://doi.org/10.1038/s41467-017-00567-8.

(15) Park, J.; Choi, S.; Yan, Y.; Yang, Y.; Luther, J. M.; Wei, S.; Parilla, P.; Zhu, K. Electronic Structure and Optical Properties of α - CH3NH3PbBr3 Perovskite Single Crystal. *J. Phys. Chem. Lett.* **2015**, *6*, 4304–4309. https://doi.org/10.1021/acs.jpclett.5b01699.

(16) DeSalvo, R.; Said, A. A.; Hagan, D. J.; Stryland, E. W. Van. Z-Scan Measurements of the Anisotropy of Nonlinear Refraction and Absorption in Crystals. *Opt. Lett.* **1993**, *18*, 194–196. https://doi.org/https://doi.org/10.1364/OL.18.000194.

(17) Walters, G.; Sutherland, B. R.; Hoogland, S.; Shi, D.; Comin, R.; Sellan, D. P.; Bakr, O. M.; Sargent, E. H. Two-Photon Absorption in Organometallic Bromide Perovskites. *ACS Nano* **2015**, *9*, 9340–9346. https://doi.org/https://doi.org/10.1021/acsnano.5b03308.

(18) Saouma, F. O.; Park, D. Y.; Kim, S. H.; Jeong, M. S.; Jang, J. I. Multiphoton Absorption Coefficients of Organic – Inorganic Lead Halide Perovskites CH3NH3PbX3 (X = Cl, Br, I) Single Crystals. *Chem. Mater.* **2017**, *29*, 6876–6882. https://doi.org/10.1021/acs.chemmater.7b02110.

(19) Sheik-Bahae, M.; Said, A. A.; Wei, T. H.; Hagan, D. J.; Van Stryland, E. W. Sensitive Measurement of Optical Nonlinearities Using a Single Beam. *IEEE J. Quantum Electron.* **1990**, *26*, 760–769. https://doi.org/10.1109/3.53394.

(20) Van Stryland, E. W.; Sheik-Bahae, M. Z-Scan Measurements of Optical Nonlinearities. In *Characterization techniques and Tabulations for Organic Nonlinear Materials, M.G. Kuzyk and C.W. Dirk, Marcel Dekker, Inc.*; 1998; pp 655–692.

(21) Sheik-Bahae, M.; Hutchings, D. C. D.; Hagan, D. J.; Van Stryland, E. W. Dispersion of Bound Electron Nonlinear Refraction in Solids. *IEEE J. Quantum Electron.* **1991**, *27*, 1296–1309. https://doi.org/10.1109/3.89946.

(22) Krishnakanth, K. N.; Seth, S.; Samanta, A.; Rao, S. V. Broadband Femtosecond Nonlinear Optical Properties of CsPbBr3 Perovskite Nanocrystals. *Opt. Lett.* **2018**, *43*, 603–606. https://doi.org/10.1364/OL.43.000603.

(23) Said, A. A.; Sheik-Bahae, M.; Hagan, D. J.; Wei, T. H.; Wang, J.; Young, J.; Van Stryland, E. W.

Determination of Bound-Electronic and Free-Carrier Nonlinearities in ZnSe, GaAs, CdTe, and ZnTe. *J. Opt. Soc. Am. B* **1992**, *9*, 405–414. https://doi.org/10.1364/JOSAB.9.000405.

(24) Saidaminov, M. I.; Abdelhady, A. L.; Murali, B.; Alarousu, E.; Burlakov, V. M.; Peng, W.; Dursun, I.; Wang, L.; He, Y.; Maculan, G.; et al. High-Quality Bulk Hybrid Perovskite Single Crystals within Minutes by Inverse Temperature Crystallization. *Nat. Commun.* **2015**, *6*, 7586. https://doi.org/10.1038/ncomms8586.

(25) Heiderhoff, R.; Haeger, T.; Pourdavoud, N.; Hu, T.; Al-khafaji, M.; Mayer, A.; Chen, Y.; Scheer, H.-C.; Riedl, T. Thermal Conductivity of Methylammonium Lead Halide Perovskite Single Crystals and Thin Films: A Comparative Study. *J. Phys. Chem. C* **2017**, *121*, 28306–28311. https://doi.org/10.1021/acs.jpcc.7b11495.

# Supporting information for "Nonlinear refraction in CH$_3$NH$_3$PbBr$_3$ single crystals"

Christian Kriso,[1,*] Markus Stein,[1] Tobias Haeger,[2] Neda Pourdavoud,[2] Marina Gerhard,[1] Arash Rahimi-Iman,[1] Thomas Riedl,[2] and Martin Koch[1]

[1]Department of Physics and Material Sciences Center, Philipps-Universität Marburg, Renthof 5, 35032 Marburg, Germany

[2]Institute of Electronic Devices, University of Wuppertal, Rainer-Gruenter-Strasse 21, 42119 Wuppertal, Germany

*christian.kriso@physik.uni-marburg.de

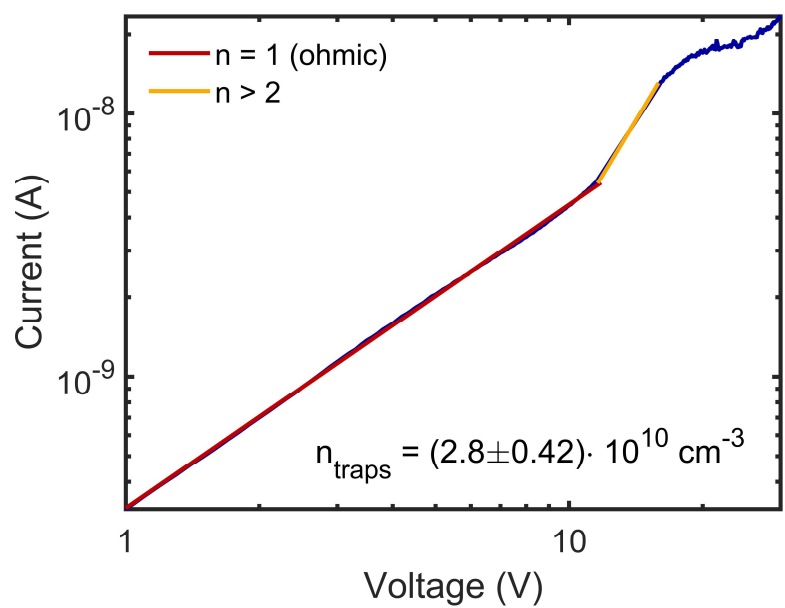

**Figure S1:** Current-Voltage traces of MAPbBr$_3$ single crystals revealing a low trap state density of $(2.8\pm0.42)\cdot10^{10}$ cm$^{-3}$. This value is nearly the same as in Saidaminov *et al.*[S1]

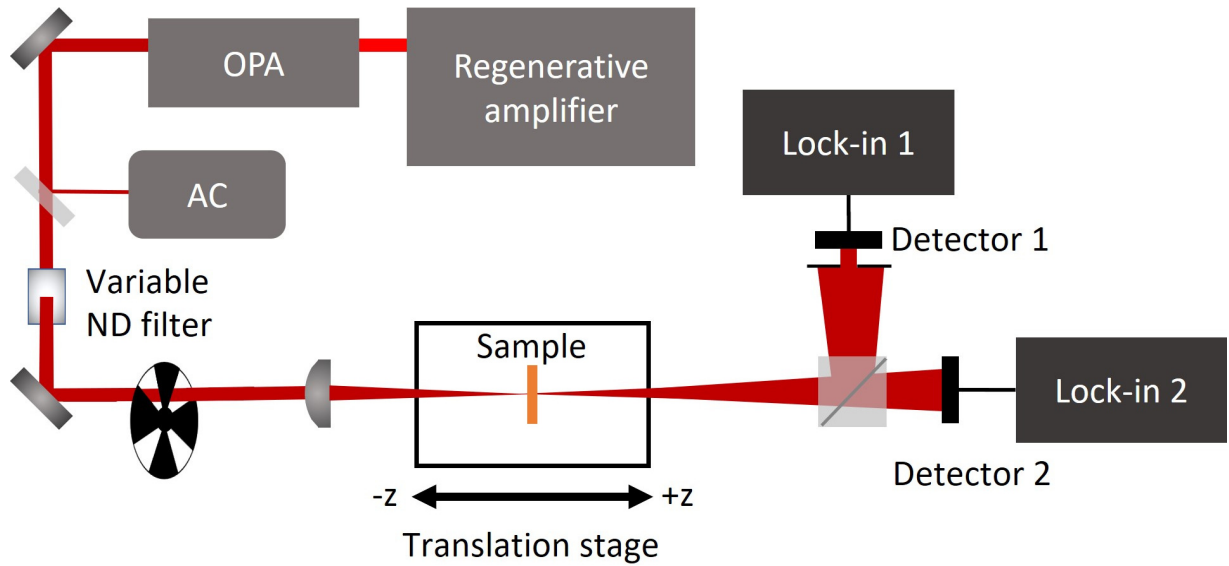

**Figure S2:** Z-scan setup for measurement of nonlinear refraction and absorption. A 1 kHz regenerative amplifier system drives an optical parametric amplifier (OPA) which emits ultrafast laser pulses centered at 1000 nm. An autocorrelator (AC) is used for pulse length determination. After passing through an ND filter with variable attenuation, the probe beam is focused by a lens with 10 cm focal length. A translation stage moves the sample through the focus of the beam, which is subsequently recorded by two large-aperture silicon photodiodes. An aperture before detector 1 only transmits the center of the beam and thus is sensitive to (de-)focusing of the beam while detector 2 records the total transmittance for the measurement of nonlinear absorption.

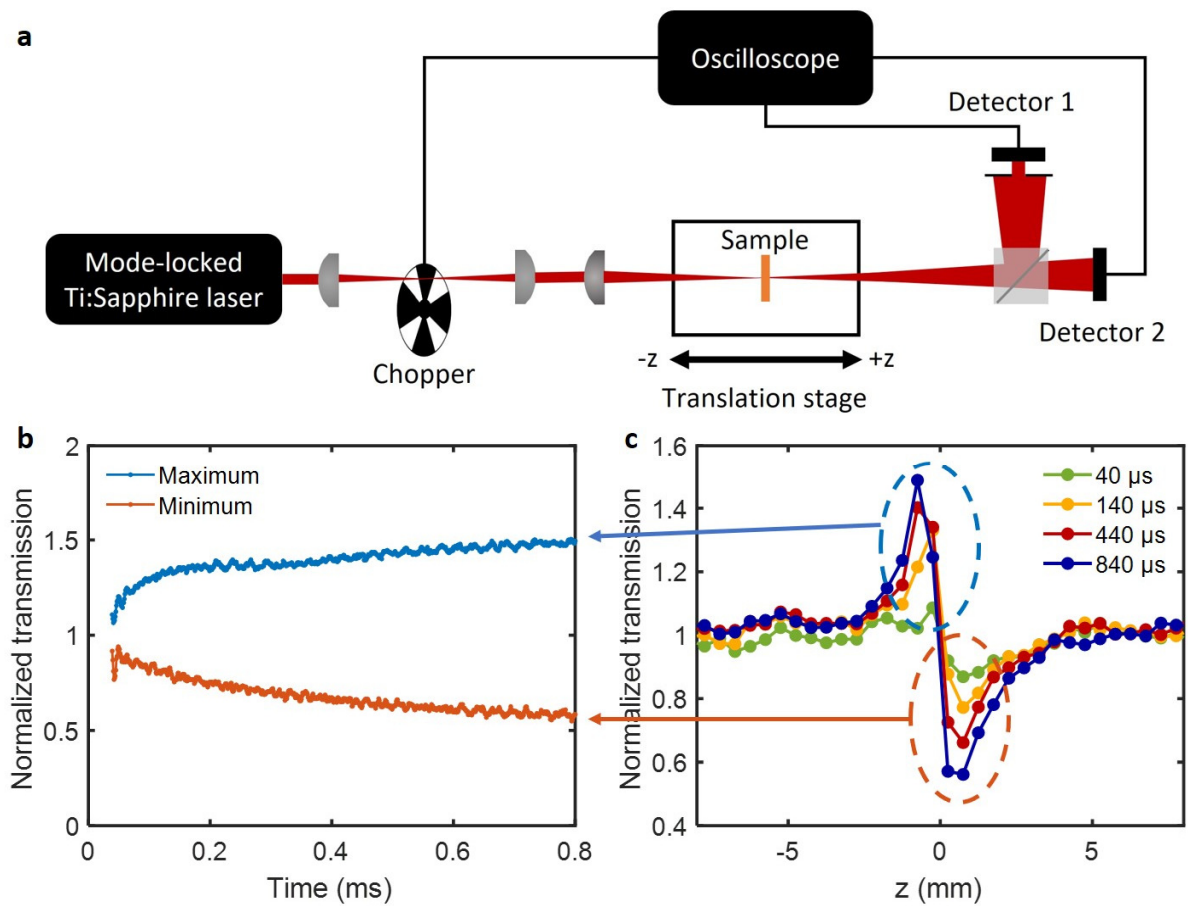

**Figure S3:** (a) Setup for evaluating the influence of a high-repetition rate laser (80 MHz) on Z-scan measurements of MAPbBr$_3$ (as done by Gnoli et al.[S2]). The opening of the chopper window triggers the oscilloscope which records the temporal development of the Z-scan signal at the minimum and maximum position which is displayed in (b). For small times (< 30μs) the signal is nearly at 0. (c) Several Z-scan measurements recorded at different times after the chopper window opens. These results demonstrate the big influence of cumulative thermal effects when using high repetition rate lasers. The measurements here have been performed at 880 nm.

## References

(S1) Saidaminov, M. I.; Abdelhady, A. L.; Murali, B.; Alarousu, E.; Burlakov, V. M.; Peng, W.; Dursun, I.; Wang, L.; He, Y.; Maculan, G.; et al. High-Quality Bulk Hybrid Perovskite Single Crystals within Minutes by Inverse Temperature Crystallization. *Nat. Commun.* **2015**, *6*, 7586. https://doi.org/10.1038/ncomms8586.

(S2) Gnoli, A.; Razzari, L.; Righini, M. Z-Scan Measurements Using High Repetition Rate Lasers: How to Manage Thermal Effects. *Opt. Express* **2005**, *13*, 7976–7981. https://doi.org/https://doi.org/10.1364/OPEX.13.007976.